# Detection of Winding Axial Deformation in Power Transformers by UWB Radar Imaging

M. S. Golsorkhi[1], M. A. Hejazi[2], G. B. Gharepetian[3], Razieh Mosayebi[3]

[1] Department of Electrical and Computer Engineering, Isfahan University of Technology, Isfahan, Iran.
[2]Electrical Engineering Department, University of Kashan, Kashan, Iran
[3]Electrical engineering department, Amirkabir university of technology, Tehran, Iran

*Abstract-* **In this paper, a novel method for detecting transformer winding axial displacement has been presented. In this method, which is based on UWB radar imaging, a UWB pulse is transmitted to the transformer winding and the reflection from it is received and recorded. By changing the antenna position this process is repeated on several points along the axis of the winding. The measured signals are mapped to a 2-D image of the winding. By analyzing this image, the axial displacement of the winding can detected and the magnitude of it is determined with an acceptable precision. Simulation results are provided to verify the proposed method.**

## NOMENCLATURE

| | |
|---|---|
| $c$ | wave speed |
| $d_M$ | distance between adjacent measuring points |
| $d_0$ | distance between two points of the image - pixel size |
| $K$ | number of measuring points |
| $M$ | number of pixels along Y-axis |
| $N$ | number of pixels along Z-axis |
| $R$ | distance between the antenna and the target |
| $Y1$ | ordinate of lower edge of the winding model |
| $Y2$ | ordinate of upper edge of the winding model |

## I. INTRODUCTION

Power transformers, which are the essential and costly parts of electric power systems, might have different electrical and mechanical faults. The transformer winding mechanical faults can be caused by mechanical forces which arise during the transformer short circuit or transportation. They can occur in radial or axial direction. In this paper, we focus on the radial deformation.

Mechanical faults can gradually weaken winding insulation capability, and cause electrical short circuit inside the transformer. This might severely damage the transformer winding and at the same time, decreases the power system reliability. However, these damages can be prevented by early detection of mechanical faults and taking the transformer out of service.

Several methods concerning mechanical fault detection have been presented in the literature. Most of these methods introduce a signature for the transformer winding. This signature is measured and registered for a particular transformer. If the signature of a transformer shows a noticeable change, it is deduced that a mechanical fault has been occurred. This signature can be the frequency response [1], the short circuit impedance [2] or the low voltage ipulse response [3] of the winding. Another method which has been presented recently, utilized a transceiver to send electromagnetic waves toward the winding and measure its reflection. The reflected wave is taken as the winding signature. This method has been implemented in the frequency [4] and time domain [5], [6].

In this paper, a novel method for an in-detail detection of the winding axial displacement is presented. This method is based on the ultrawide-band (UWB) radar imaging, which has been used for a wide range of applications [7] ,[8].

In UWB radar imaging method, a pulse transceiver is used to transmit a short pulse toward the target, and receive and the waves reflected from it. The measuring process is repeated at several points. The set of measured signals are further are mapped to an image of the target. This image reveals geometrical and electromechanical specifications of the target [9].

In this paper, UWB radar imaging is utilized to take a 2-D picture of transformer winding. This picture is in the vertical direction; consequently the winding appears as a rectangle in it. If axial displacement occurs in the winding, the location of the upper or lower edge of this rectangle will be changed in the resultant image. Therefore this edge is considered as a reference to detect the occurrence of the axial displacement as well as the magnitude of it.

In order to verify the proposed method, the transformer winding and the transceiver have been modeled in CST software, and the UWB imaging procedure is simulated. The simulations results are further processed to take a 2-D image of the winding. This process is conducted with and without axial displacement. The resultant UWB images are compared to determine the magnitude of the axial displacement.

## II. UWB IMAGING

### A. Data acquisition

The principle of UWB radar imaging is shown in Fig. 1. An UWB transceiver is utilized to generate a short pulse, and transmits the pulse through the antenna toward the target. The transmitted pulse hits the target and a fraction of its energy reflects back to the antenna. The received signal is a UWB pulse, which has a time delay proportional to the distance between the antenna and the target. In other words, we have:

$$T_d = 2R/c \qquad (1)$$

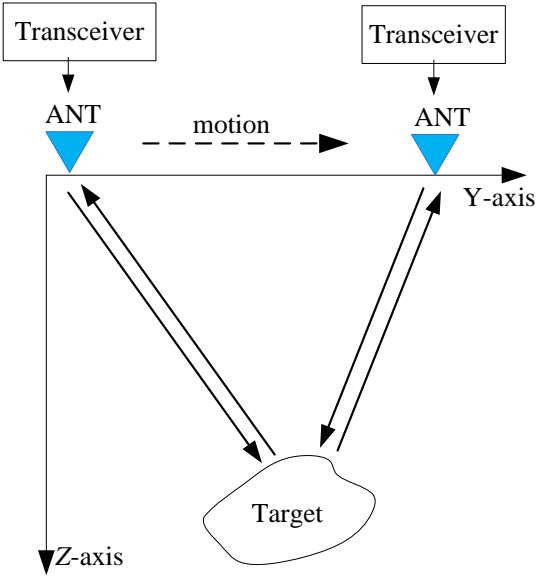

Fig. 1. UWB radar imaging

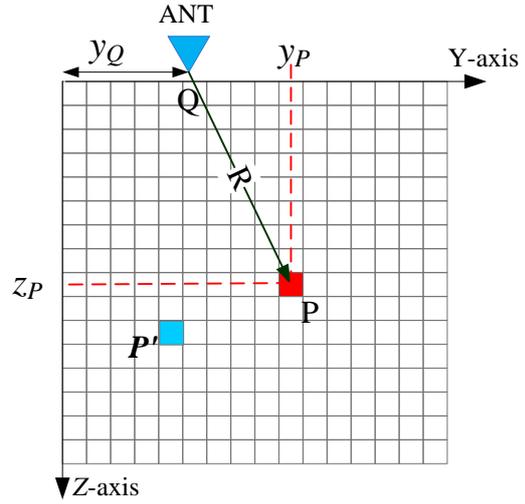

Fig.2. Coordinates of TA migration

Now, by changing the position of the antenna, the target reflection is measured on different points of Y-axis. Note that since the antenna's beam angle is wide, the target can be "seen" from a wide range of points along Y-axis. The distance between adjacent measuring points is a constant and is defined as measuring step. Each measured signal is a discrete-time function, and can be represented as a vector. The set of measured signals constitute a 2-D matrix, which is a function of time ($t$) and antenna position. It forms the basic data for radar imaging. In section 2-c, it will be shown that this 2-D matrix can be mapped to 2-D image matrix.

### B. Data preprocessing

In section A, only one object is assumed to exist in the environment, i.e., the target. However, in case of this paper, the environment is the transformer tank and the target is the transformer winding. Therefore, in addition to the target, there are various objects in the environment, none of which are of interest. Therefore each measured signal consists of several short pulses related to different objects. Since the distances of these objects from the antenna are different, the time delay of the corresponding received pulses is also different. The unwanted reflections can be removed, by preserving the section of the signal which includes the target reflection and removing other parts of it. Since the target, i.e., transformer winding is stationary, for a specific measuring point, the time delay of the target reflection has a specific value. However, since the transformer winding is a 3-D object, there are several paths between antenna and the winding. The shortest path is related to the nearest point of the winding to the antenna and the longest path is related to the farthest point of it. Therefore, the winding reflection lies in the time interval of ($T_{min}$,$T_{max}$), where $T_{min}$ and $T_{max}$ are the time delay related to the shortest and longest distance between winding and antenna, respectively. This interval should be calculated for each measuring point, and the section which lies in this interval should be extracted from the measured signal [12].

### C. Migration algorithm

The process of obtaining an image from the matrix of measured signals is called migration. Several migration algorithms have been presented in the literature [10]. In this paper, the time arrival (TA) migration algorithm is utilized [11]. It is based on the wave travel time from the antenna to the target and back to the antenna.

The principle of time arrival algorithm is shown in Fig. 2. Suppose that there is a point target at point P($y_P$, $z_P$) on the Y-Z plane, and the antenna is located at the measuring point Q($y_Q$, 0) on the Y-axis. The time delay of the signal measured at the $k^{th}$ measuring point ($kd_M$) is calculated, as follows[13]:

$$T_{dP,k} = \frac{2}{c}\sqrt{\left(y_P - kd_M\right)^2 + z_P^2} \qquad (2)$$

As the antenna is moved along the Y-axis, $T_{dP,k}$ varies as shown in Fig. 3. Now, we calculate $T_{dP,k}$ by applying (2) at each measuring point, shift back the corresponding signal at that point, and sum the shifted signals, as follows [14]:

$$J(y_p, z_p, t) = \sum_k b_k\left(t + T_{dP,k}\right) \qquad (3)$$

where, $b_k$ is the signal measured at the $k^{th}$ measuring point. $J$ is a time signal, which consists of a short pulse at t=0, as shown in Fig. 3. Note that since the target is located at point P, the calculated time delays ($T_{dP,k}$) match the actual time delays of the measured signals, and as a result, $J$ has a great value at $t$=0. If we calculate $J$ for another other point of the Y-Z plane, for example $P'$, $T_{dP',k}$ will not match the time delay of the measured signals, and the value of $J$ at t=0 will either be small or zero. Therefore, the value of $J(y,z,0)$ at any point can be regarded as a criterion that indicates the existence of target at that point. In other words, if $J(y,z,0)$ is plotted over the Y-Z plane, the result would be an image of the target. We define the image function, $I$, as follows:

$$I = J(y,z,0) = \sum_k b_k\left(T_{dP,k}\right) \qquad (4)$$

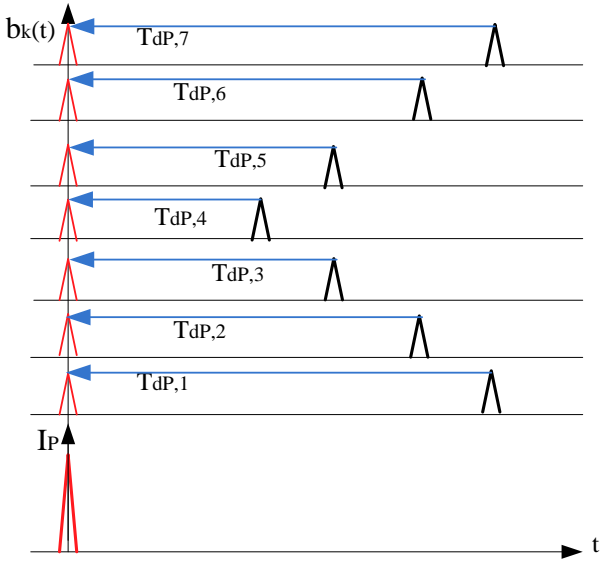

Fig. 3. Principle of TA migration

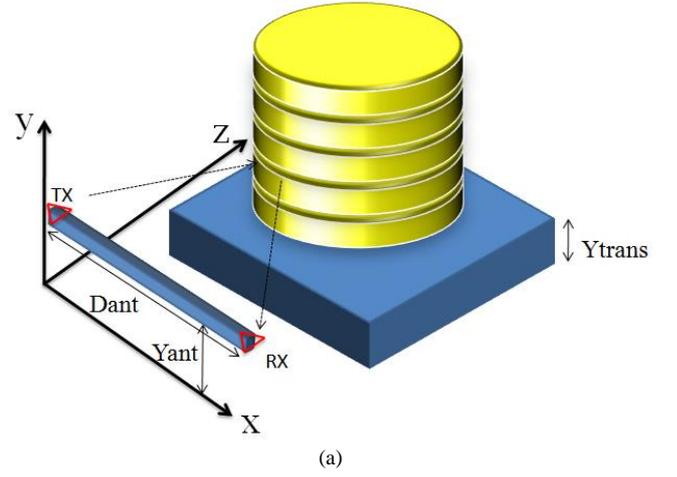

(a)

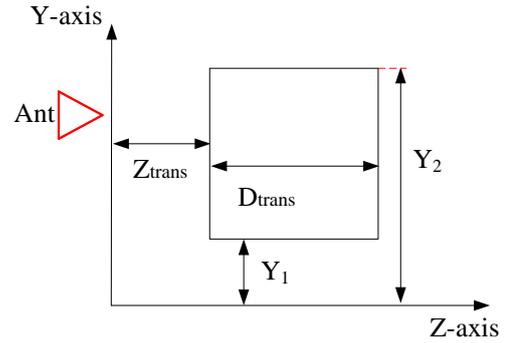

(b)

Fig. 4. Transformer winding model: a) 3-D view and b) 2-D side view

The UWB radar imaging process is summarized below:

Step 1. Perform measurement at several points of Y-axis

Step 2. calculate the time window of target reflection for each measuring point

Step 3. Extract the target reflection for each measuring point ($b_k(t)$)

Step 4. Initialize parameters: $i$=0, $j$=0, $k$=0, $d_0$, $d_M$, $K$, $M$, $N$ and I=$[0]_{M*N}$

Step 5. Calculate the wave propagation time from the kth measuring point to the pixel (i,j), as follows:

$$T_{d,k} = \frac{2}{c}\sqrt{\left(id_0 - kd_M\right)^2 + \left(jd_0\right)^2} \quad (5)$$

Step 6. Calculate the I at pixel (i,j) by using (4)

Step 7. Increment k by 1. If $k<K$ go to the step 3, otherwise go to the next step.

Step 8. Increment i by 1. If $i<M$ go to the step 3, otherwise go to the next step.

Step 9. Increment j by 1. If $j<N$ go to the step 3, otherwise go to the next step.

Step 10. Plot I

## III. SIMULATION RESULTS

In this paper we utilize the UWB imaging procedure to detect the axial deformation of transformer winding. Note that since the transformer tank is made of conductor materials, electromagnetic waves can't penetrate through it. Therefore, in order to implement UWB imaging in an actual transformer, we need to create a window in the transformer tank by cutting some part of it and filling it with an insulator. Then we place the antennas in front of the insulator window.

The transmitted UWB pulse passes through the insulator window, reaches the winding, and reflects back to the antenna.

In order to detect the axial displacement of the transformer winding, we need to create a 2-D SAR image in the vertical direction. This image lies in the Y-Z plane, where Y and Z axes are assumed to be parallel and perpendicular to the winding axis, respectively. Therefore the measuring process is conducted by moving the antennas along a vertical line on the surface of the transformer tank.

In this paper, the UWB imaging process is simulated on CST software. As shown in Fig. 4.a, the transformer winding is modeled by 6 disks, covered by copper and located at the top of each other. The axial displacement of the winding is modeled by moving the winding model in the vertical direction. Two Vivaldi type antennas are modeled and placed in front of the winding model. For each measuring point, the antennas height is changed and the simulation process is repeated. The received signals are saved and used as raw data for the migration algorithm.

The side view of the model is illustrated in Fig.4.b. In this Figure the model is shown as a rectangle and the ordinate of its lower and upper side are shown by $Y_1$ and $Y_2$, respectively. The resultant SAR image of the model is expected to match this Figure.

The specifications of the simulation are listed table 2. In order to assess the performance of the proposed method, two simulations are performed. In the first simulation, it is assumed that the winding is intact, i.e., without axial displacement. In the second simulation, an axial displacement of 20mm is introduced in the winding, by changing the position of the winding in the vertical direction.

Table 1. The specification of the simulation

| Parameter | Symbol | Value (mm) |
|-----------|--------|------------|
| Model cylinder height | - | 150 |
| Model diameter | - | 300 |
| Model height (intact winding) | $Y_1$ | 230 |
| Model height (displaced winding) | $Y_2$ | 250 |
| Model distance from the Y-axis | $Z_{mdl}$ | 450 |
| Measuring step | $d_M$ | 30 |
| Number of measuring points | K | 21 |

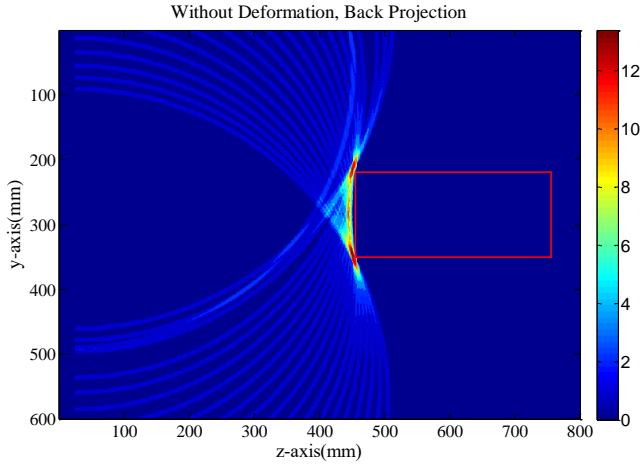

Fig. 5. UWB images of the model for the first simulation

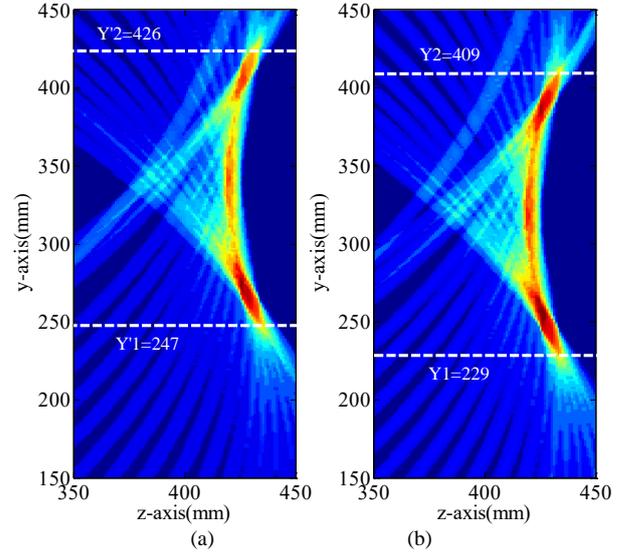

Fig. 6. Zoomed UWB images of the model in: a)first simulation and b)second simulation

Table 2. The summery of experimental results

| Parameter | Sym. | Descript. | Simulation 1 | Simulation 2 |
|-----------|------|-----------|--------------|--------------|
| Lower edge ordinate (mm) | $Y_1$ | measured | 247 | 229 |
| Upper edge ordinate (mm) | $Y_2$ | measured | 426 | 409 |
| Center ordinate (mm) | $Y_C$ | estimated | 336.5 | 319 |
| Axial displacement(mm) | AD | estimated | - | 17.5 |
| Axial displacement(mm) | AD | actual value | - | 20 |
| Estimation error | - | - | - | 12.5% |

Fig. 5 shows the UWB image of the model in the first simulation. In this picture, the image function is shown in color-map scale, i.e., the magnitude of the image function at any point is shown by an appropriate color. The red and yellow colors indicate a high and moderate reflection, respectively. The actual position of the model is illustrated by a red rectangle. Since, most of the transmitted signal is caught by the left side of the model, only this part of the winding has appeared as a line in the image. Although this vertical line does not match the shape of the model, its location indicates the position of the model.

Fig. 6.a and b depict the zoomed UWB image of the model corresponding to the first and second simulations, respectively. In each figure, two white horizontal lines indicate the upper and lower edges of the model.

The ordinates of the lower and upper edge of the model in each case are summarized in table 3. In addition, the ordinate of the center of the model is calculated as:

$$Y_C = \frac{Y_1 + Y_2}{2} \qquad (14)$$

We take $Y_C$ as a reference for measuring the ordinate of the model. Then we calculate the axial displacement of the model by subtracting $Y_C$ in the second simulation from that of the first simulation. Finally, the error of the proposed method in is calculated by comparing the measured value of the axial deformation with its actual value. As shown in table 3, the error is 25%. Therefore the presented method is capable of estimating the magnitude of winding axial displacement with an acceptable precision.

## IV. CONCLUSION

A novel method for detecting and measuring the magnitude of the transformer winding axial deformation based on UWB imaging has been presented. In order to verify the proposed method, the transformer winding has been modeled the CST software. Simulation results have been further processed to form a 2-D image of the model with and without axial displacement. By comparing them, In addition to detecting the occurrence of axial displacement, the magnitude of it is determined with an acceptable precision.